\documentclass[aps,prd,amsfonts,nofootinbib,supercriptaddress]{revtex4}
\usepackage{amssymb,amsmath}
\usepackage{graphicx}
\usepackage{epsfig}

\newcounter{fig}

\newcommand{\beq}{\begin{equation}}
\newcommand{\eeq}{\end{equation}}
\newcommand{\bea}{\begin{eqnarray}}
\newcommand{\eea}{\end{eqnarray}}

\begin{document}

\hfill Preprint numbers: ITP-UU-12/32,  SPIN-12/30
%FER-xx-xx
%arXiv:1209.xxxx
%\leftline{\hfill Date: \today}

%\vspace{0.5cm}

\title{Antiscreening in perturbative quantum gravity and resolving the Newtonian singularity}

\author{Anja Marunovi\'c$^{a*}$}
\author{Tomislav Prokopec$^{b}$}
\email[]{anja.marunovic@fer.hr, t.prokopec@uu.nl}
%\author{Anja Marunovi\'c$^{a}\footnote{anja.marunovic@fer.hr}$}
%\author{Tomislav Prokopec$^{b}\footnote{t.prokopec@uu.nl}$}
%\email[]{anja.marunovic@fer.hr, t.prokopec@uu.nl}
\affiliation{$^a$University of Zagreb, Faculty of Electrical
Engineering and Computing, Physics Department, Unska 3, 10000 Zagreb, Croatia\\
 $^b$Institute for Theoretical Physics and Spinoza
Institute, Utrecht University, Leuvenlaan 4, 3584 CE Utrecht, The
Netherlands }

\begin{abstract} \noindent

 We calculate the quantum corrections to the Newtonian potential induced
by a massless, nonminimally coupled scalar field on Minkowski
background. We make use of the graviton vacuum polarization
calculated in our previous work and solve the equation of motion
nonperturbatively. When written as the quantum-corrected
gauge invariant Bardeen potentials, our results show
that quantum effects generically antiscreen the Newtonian
singularity $\propto 1/r$. This result supports the point of view
that gravity on (super-)Planckian scales is an asymptotically safe theory.
In addition, we show that,
in the presence of quantum fluctuations of a massless,
(non)minimally coupled scalar field,
dynamical gravitons propagate superluminally.
The effect is, however, unobservably small and it is hence of academic
interest only.

\end{abstract}

%PACS:
%04.62.+v Quantum fields in curved spacetime
%04.70.Dy Quantum aspects of black holes, evaporation, thermodynamics

\maketitle

\section{Introduction}

In Ref.~\cite{Marunovic:2011zw} we considered
the graviton vacuum polarization induced by the one-loop quantum fluctuations
of a nonminimally coupled, massless scalar field on Minkowski background in
the Schwinger-Keldysh formalism. This allowed us to compute
the large distance, leading order, quantum correction to
the scalar gravitational (Bardeen) potentials,
which in longitudinal gauge reduce to the Newtonian potentials.
We used the Schwinger-Keldysh formalism in direct (coordinate) space
which allowed us to study {\it time transients}
associated with switching on scalar vacuum fluctuations
at a definite moment in time. Such transients can be induced, for example,
by symmetry restoration that renders the scalar field massless.
Moreover, just as in the presence of classical matter,
the two quantum Bardeen potentials differ when the effects
of quantum (scalar) matter fluctuations are included.

In Ref.~\cite{Marunovic:2011zw} we adopted the effective approach to
quantum gravity (see {\it e.g.} Refs.~\cite{Donoghue:1994A,Donoghue:1994B}),
according to which quantum gravitational effects can be reliably calculated
by perturbative methods,
provided one focuses on low energy and momentum phenomena.\footnote{
By low energy/momentum phenomena we mean the phenomena whose energies
$E$ and momenta $p=\|\vec p\,\|$ are well below the Planck scale,
$E\ll E_P$, $p\ll E_P/c$, where
$E_P=m_Pc^2=\sqrt{\hbar c^5/G_N}\simeq 1.2\times 10^{19}~{\rm GeV}$,
where $m_P$ is the Planck mass,
$c$ is the speed of light, $G_N$ denotes the Newton constant and
$\hbar = 1.055\times 10^{-34}~{\rm Js}$ is the reduced Planck constant.}
By now, there is quite a long history and extensive literature
on long range, quantum corrections to the Newtonian potentials.
Such studies were pioneered by
Donoghue~\cite{Donoghue:1994A,Donoghue:1994B},
and followed by many others~\cite{HamberLiu:1995,Akhundov:1997},
with differing results. The problem was again revived in the 2000s in
Refs.~\cite{BjerrumBohr:2002A,BjerrumBohr:2002B,BjerrumBohr:2003C,Butt:2006,Faller:2008,Dalvit:1994gf,Satz:2004hf},
where also some of the graviton vertex corrections were included.
The Schwinger-Keldysh formalism was for the first time used
in Ref.~\cite{Park:2011ww}. Related older works
include~\cite{Capper:1979ej,Capper:1978yf}
where the one-loop graviton vacuum polarization induced by
photons on a Minkowski background was calculated.

 Here we extend our work in~\cite{Marunovic:2011zw} and present a resummation
scheme by which we consider short and large scale screening effects
of gravitational potentials induced by quantum fluctuations of
a massless nonminimally coupled scalar.
More precisely, we make use of the one-particle irreducible (1PI) resummation
technique and arrive at the result that the Newtonian potentials exhibit
a strong {\it antiscreening} at length scales of the order of
the Planck scale. Surprisingly, there are also long range oscillatory quantum
corrections to the Newtonian potentials. Even though these corrections
are present at large distances, they are
high energy corrections, since oscillations occur at the Planck length
scale.

 Admittedly, these results do not belong to the realm of
effective quantum gravity, questioning their reliability.
Still, we feel compelled to present them, and believe that they deserve some
attention. Indeed, this work is not the first attempt
to use a resummation technique to investigate short scale effects
in quantum gravity. The main other contender is known as
the {\it effective action} approach
to quantum gravity~\cite{Reuter:1996cp}, and it is usually implemented within
the {\it asymptotic safety} program for quantum gravity,
and we shall now recall the basic idea behind this approach.

 The asymptotic safety program in quantum gravity was initiated by
 Weinberg~\cite{Hawking:1979ig}, where he argued
that a weaker requirement than the usual renormalizability
may be imposed on quantum gravity.
Recently, this program has attracted a considerable attention
among researchers in quantum gravity~\cite{Reuter:2012id,Niedermaier:2006ns}.
If quantum gravity were an asymptotically safe theory, it would exhibit
few-dimensional critical hypersurface characterizing
the running of relevant couplings
(that are associated with relevant operators).
The running of these couplings "emanate" from an ultraviolet (UV) fixed point,
at which their value has a definite value that can be, in principle,
determined. The claim is that the set of relevant operators
consists of unity (the corresponding coupling being the cosmological term),
the Ricci scalar (the corresponding coupling being the Newton constant) and
possibly some quadratic curvature invariants. There is limited evidence
that all other (higher derivative) operators are irrelevant, in the sense
that the UV fixed point "repels" the corresponding couplings
as one runs towards the ultraviolet. The value of these irrelevant couplings
is (at least in principle) uniquely determined by the requirement that
the physical theory runs towards the UV fixed point, as the running scale
approaches infinity. Thus, an asymptotically safe quantum gravity consists of
several (relevant) couplings whose value ought to be determined
by measurements, for example in the far infrared, while
all other coupling parameters are given by the requirement
of asymptotic safety. Within this framework it is not true
anymore that there are infinitely many counterterms appearing
in perturbation theory, whose coefficients (couplings) must be determined
by (an infinite number of) measurements. In this sense
the asymptotic safety program resolves the problem of nonrenormalizability
of quantum gravity. What remains to be shown is
that quantum gravity is indeed an asymptotically safe theory, which is far from
a trivial pursuit.

 Indeed, the proof that quantum gravity is asymptotically safe
would require solving quantum gravity nonperturbatively,
which is very hard. Instead,
researchers have focused on developing perturbative and nonperturbative
methods (for reviews, see~\cite{Reuter:2012id,Niedermaier:2006ns}),
hoping to collect enough evidence in support of the asymptotic
safety hypothesis.
The original evidence due to Weinberg~\cite{Hawking:1979ig}
is based on an $\epsilon$-expansion around two spacetime dimensions.
Here, $\epsilon=2$ is not a small number in four spacetime
dimensions, making the whole approach questionable.
The second approach is based on a functional renormalization
group equation for the gravitational effective action pioneered by
Reuter~\cite{Reuter:1996cp}. While the starting
equation is an exact (functional) differential equation for the effective
gravitational action, the actual approximation scheme is limited
to an {\it a priori} chosen set of local operators,
truncated at some order (usually fourth) in
space and time derivatives. The criticism of the method is twofold.
Firstly, the operators are chosen {\it a priori} based on
a derivative expansion which is essentially a low energy expansion,
while the application is extended to the ultraviolet sector.
Secondly, the operators of canonical dimension four or
higher generically violate unitarity,
questioning thus the validity of the whole (nonperturbative) scheme.
Yet, the results (if true) are fascinating:
they suggest existence of an ultraviolet (UV)
non-Gaussian, nontrivial, fixed point in quantum gravity.
Close to this UV fixed point the Newton constant runs with
a scale $\mu$
as  $G_N(\mu) \simeq g_*/\mu^2$ (as $\mu\rightarrow \infty$) ,
while the cosmological constant
runs as $\Lambda(\mu) \simeq \lambda_*\mu^2$, where
$g_*\simeq 0.27$ and $\lambda_*\simeq 0.35$ are calculable (nonvanishing)
real constants.
While the traditional approach to asymptotic safety is to
demand that one obtains the Einstein-Hilbert action in
the deep infrared, recently Satz \emph{et at}.~\cite{Satz:2010uu} have developed an effective average
action approach starting in the infrared with the Barvinsky-Vilkovisky
non-local effective action~\cite{Barvinsky:1987uw,Barvinsky:1990up},
which takes account of one-loop massless field excitations of
fields with various spins.

 In this work we present a different resummation scheme,
such that antiscreening occurs as a result of resumming
the one-loop 1PI diagrams for the gravitational
potentials. In contrast to the local operator approach mentioned above,
where a derivative expansion is used,
our resummation is based on a nonlocal correction to the graviton vacuum
polarization, and thus does not suffer from the above problems.
The hope is that, if we choose the coefficients of the finite counterterms
to be zero, our method will not violate unitarity.
Furthermore, our method does not violate any of the symmetries of the theory.
Namely, we linearize in the gravitational field, and  in our analysis we
maintain the linear version of diffeomorphism invariance
(also known as gauge invariance).

 The paper is organized as follows: in Sec. II we present
a self-consistent solution to the one-loop 1PI effective field equations
for the gravitational potentials.
In Sec. III we discuss our results and give an outlook for future work.

Unless stated explicitly, we work in natural units where
$\hbar=1=c$.

\section{Resummation}

The quantum corrected equation of motion for the linearized Einstein equation
(see Eq.~(39) in~\cite{Marunovic:2011zw}) is of the form
\beq
L_{\mu\nu\rho\sigma}h^{\rho\sigma}(x) +\int d^4
x^\prime\left[{}_{\mu\nu}\Pi_{\rho\sigma}^{\rm ret}\right]
       (x;x^\prime)h^{\rho\sigma}(x^\prime) + {\cal O}((h_{\mu\nu})^2)
=\frac{\kappa^2}{2}\delta_\mu^0\delta_\nu^0 M\delta^3(\vec x),
\label{Full EOM}
\eeq
where $\left[{}_{\mu\nu}\Pi_{\rho\sigma}^{\rm
ret}\right](x;x^\prime)$ is the retarded graviton vacuum
polarization tensor of the graviton field perturbation
$h_{\mu\nu}(t,\vec{r}\,)$, $M$ is a (static) particle's mass and
$\kappa^2=16\pi G_N$; $G_N$ is Newton's constant.
A massless, nonminimally coupled scalar field on Minkowski background
induces at one-loop the following renormalized graviton vacuum polarization
tensor:
\bea
%&&
\left[{}_{\mu\nu}\Pi_{\rho\sigma}^{\rm ret}\right](x;x^\prime)
  &=& -\frac{\kappa^2}{7680\pi^3}\times
L_{\mu\nu\alpha\beta}
\bigg\{D^{\alpha\beta}{}_{\rho\sigma}-\eta^{\alpha\beta}D_{\rho\sigma}
  \bigg[\frac{1}{6}+30\Big(\xi-\frac{1}{6}\Big)^2\bigg]\bigg\}
\nonumber\\
 &&\hskip 1.4cm
\times\,\partial^4\,
 \Theta(\Delta t^2-\Delta r^2)\Theta(\Delta
t)\left[\ln| \mu^2(\Delta r^2-\Delta t^2)|-1\right]
\,,
\label{retarded self energy}
\eea
where
$\Delta r=\|\vec x-\vec x^{\,\prime}\|$, $\Delta t=t-t^\prime$
($\vec x$, $\vec x^{\,\prime}$, $t$ and $t^\prime$ denote spatial and
time coordinates, respectively), $\xi$
measures the coupling strength of the scalar filed $\varphi$ to
gravity through the Ricci scalar $R$,~\footnote{The covariant action of the
nonminimally coupled, massless scalar field $\varphi$ is
\begin{equation*}
S_\varphi=\int d^D x \sqrt{-g}\Big\{-\frac 12
(\partial_\mu\varphi)(\partial_\nu\varphi)g^{\mu\nu}-\frac 12 \xi
R\varphi^2\Big\}
\,.
\end{equation*}
}
and $L_{\mu\nu\rho\sigma}$ denotes the Lichnerowicz operator on Minkowski
background\footnote{Recall that, up to boundary terms,
the quadratic Einstein-Hilbert
action can be written in terms of the Lichnerowicz operator as
\begin{equation*} S_{EH}=\frac{1}{2\kappa^2}\int d^Dx
         \Big[h^{\mu\nu}L_{\mu\nu\rho\sigma}h^{\rho\sigma}
           + {\cal O}(h_{\mu\nu}^3)
         \Big]
\,.
\end{equation*}}:
\beq L_{\mu\nu\rho\sigma} = D_{\mu\nu\rho\sigma}
                     - \frac{1}{2}\eta_{\mu\nu}D_{\rho\sigma}\,,
\label{Lichnerowicz}
\eeq
with the explicit form of the operators,
\beq
D_{\rho\sigma}=\partial_{\rho}\partial_\sigma-\eta_{\rho\sigma}\partial^2
\;,\qquad D_{\mu\nu\rho\sigma} =
\partial_{(\rho}\eta_{\sigma)(\mu}\partial_{\nu)}
         - \frac{1}{2}\eta_{\mu(\rho}\eta_{\sigma)\nu}\partial^2
         - \frac{1}{2}\eta_{\rho\sigma}\partial_\mu\partial_\nu
\,,
\label{D1:D2}
\eeq
$\partial^2=\eta^{\gamma\delta}\partial_\gamma\partial_\delta$
and indices in~(\ref{Full EOM}) are lifted with the Minkowski metric,
$\eta^{\mu\nu}={\rm diag}(-1,1,1,1)$. Analogous results were found earlier
in~\cite{Horowitz:1980fj,Jordan:1987wd},
albeit the equation of motion was presented in a form somewhat
different from~(\ref{Full EOM}) and~(\ref{retarded self energy}).

\begin{figure}
\begin{center}
\leavevmode
\includegraphics[scale=1.2]{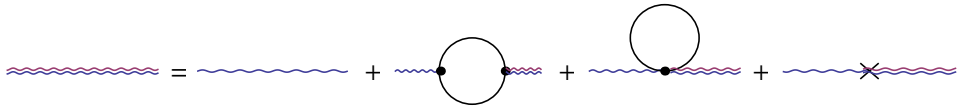}
\end{center}
\caption{The resummed one-loop contribution from scalar vacuum fluctuations to the graviton propagator.} \label{feynman1}
\end{figure}

\begin{figure}
\begin{center}
\leavevmode
\includegraphics[scale=1.2]{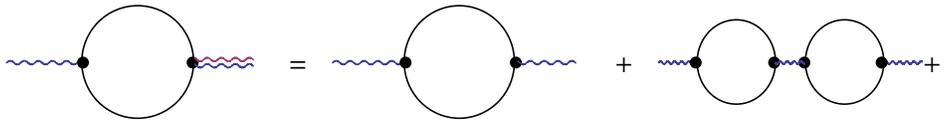}
\end{center}
\caption{The resummed contribution from the nonlocal (cubic) vertices to the graviton propagator.
The resummed loop can be written as a sum of the one-loop, two-loop, {\it etc.}, contributions.} \label{feynman2}
\end{figure}

In this paper we solve the 1PI equation of motion for the graviton one-point function~(\ref{Full EOM})
exactly. This corresponds to resumming all 1PI one-loop diagrams,
illustrated in Fig.~\ref{feynman1}, where we show the nonlocal contribution arising from cubic graviton-scalar vertices,
the local contribution from quartic scalar-graviton vertices, and the counterterm contribution.
In order to illustrate how the resummation works, in Fig.~\ref{feynman2} we show
the resummed nonlocal diagrams (arising from cubic vertices).
In principle, the contribution from the local diagrams needs to be resummed as well. But, in
a flat spacetime background they contribute only as power-law divergences, their contribution
gets automatically subtracted in dimensional regularization,
and thus they do not contribute~\cite{Marunovic:2011zw}.

To proceed, note first that it helps to insert
instead of the source on the right-hand side of~(\ref{Full EOM})
the classical equation of motion
\beq
L_{\mu\nu\rho\sigma}h^{\rho\sigma(c)}(x)=\;\frac{\kappa^2}{2}\;\delta_\mu^0\delta_\nu^0
M\delta^{D-1}(\vec x),
 \label{EOM cl}
 \eeq
where superscript $c$ stands for \emph{classical}.
This enables us to completely remove the Lichnerowicz operator
from Eq.~(\ref{Full EOM}) and thus relate the quantum gravitational field
to the classical potential through an integral
equation\footnote{In general,
the classical solution $h^{\alpha\beta(c)}$ in
Eq.~(\ref{quantum in terms of cl}) includes also gravitational
waves, which are solutions to the homogeneous equation,
$L_{\mu\nu\rho\sigma}h^{\rho\sigma(gw)}(x)=0$.
They are not of main interest here, since
we are mainly interested in the gravitational response to a static
pointlike massive particle.
}:
\bea
h^{\alpha\beta(c)}(x)&=&h^{\alpha\beta}(x)-\frac{\kappa^2}{7680\pi^3}
\bigg\{D^{\alpha\beta}{}_{\rho\sigma}-\eta^{\alpha\beta}D_{\rho\sigma}
  \bigg[\frac{1}{6}+30\Big(\xi-\frac{1}{6}\Big)^2\bigg]\bigg\}
  \nonumber \\
    &&
\times\partial^4\int
d^4 x^\prime\Theta(\Delta t^2-\Delta r^2)\Theta(\Delta
t)\left[\ln(\mu^2(\Delta t^2-\Delta
r^2))-1\right]h^{\rho\sigma}(x^\prime).
\label{quantum in terms of cl}
 \eea
This form of the equation suggests that the vacuum polarization
can be thought of as a (nonlocal, spacetime dependent, causal)
wave function renormalization of the graviton $h^{\alpha\beta}(x)$.
In the Appendix we show how to solve this equation.
The general solutions in momentum space are given in
Eqs.~(\ref{A:Phi:3}) and~(\ref{A:Psi:3}). Here we are interested in a static response
to a point mass, for which the classical Bardeen potentials
in momentum space are $\Phi^{(c)}=\Psi^{(c)}=-[4\pi G_NM/k^2]\times 2\pi\delta(k^0)$,
$k\equiv \|\vec k\|$
(see Ref.~\cite{Marunovic:2011zw}). Hence, a static response to a point mass $M$
is obtained simply by taking the limit $k^0\rightarrow 0$
of Eqs.~(\ref{A:Phi:3}) and~(\ref{A:Psi:3}):
\begin{eqnarray}
\Phi(k^\mu) &=& -\frac{4\pi G_NM\times 2\pi\delta(k^0)}{k^2}\Bigg[
                \frac43\frac{1}{1+k^2{\cal K}}
                -\frac13\frac{1}{1-5(1-6\xi)^2k^2{\cal K}}
             \Bigg]\,,
\qquad
\label{Phi:3}\\
\Psi(k^\mu) &=& -\frac{4\pi G_NM\times 2\pi\delta(k^0)}{k^2}\Bigg[
                \frac13\frac{1}{1+k^2{\cal K}}
                +\frac23\frac{1}{1-5(1-6\xi)^2k^2{\cal K}}
             \Bigg]
\,,
\label{Psi:3}
 \end{eqnarray}
where, in the static limit,
\begin{equation}
k^2{\cal K} =  \frac{G_Nk^2}{60\pi}
           \Big[\ln\Big(\frac{k}{2\mu}\Big)+\gamma_E \Big]
\,.
\label{cal K}
\end{equation}
Since for the vector and tensor perturbations
classical fields vanish, Eqs.~(\ref{A:ni:2}) and~(\ref{A:hijTT:2}) imply that,
at one-loop order, also the quantum corrected fields vanish,
     $\tilde n^T_i=0=h_{ij}^{TT}$.

Upon performing an inverse Fourier transformation and setting
$x\equiv k l_{Pl}$, $\tilde\mu\equiv\mu l_{Pl}{\rm e}^{-\gamma_E}$, $y\equiv
r/l_{Pl}$ (where $l_{Pl}=\sqrt{G_N}$ is the Planck length), we
arrive at the solution for the gauge invariant scalars in position space:
\bea
\Phi(r)&=&-\frac{2G_N M}{\pi r}\int_0^{\infty}\frac{dx \sin(x y)}{x}
  \left\{\frac 4 3\frac{1}{1+\frac{x^2}{120\pi}
             \ln\left(\frac{x^2}{4\tilde\mu^2}\right)}
-\frac13\frac{1}{1-(1-6\xi)^2\frac{x^2}{24\pi}
     \ln\left(\frac{x^2}{4\tilde\mu^2}\right)}
       \right\}\,,
\label{Phi:4}
\\
\Psi(r)&=&-\frac{2G_N M}{\pi
r}\int_0^{\infty}\frac{dx \sin(x y)}{x}\left\{\frac 2 3
\frac{1}{1+\frac{x^2}{120\pi}\ln\left(\frac{x^2}{4\tilde\mu^2}\right)}
+\frac 1 3
\frac{1}{1-(1-6\xi)^2\frac{x^2}{24\pi}\ln\left(\frac{x^2}{4\tilde\mu^2}\right)}
\right\}\,.
\label{Psi:4}
 \eea
As such these integrals cannot be computed analytically. However,
it is feasible to examine their behavior for small $r$ and for
large $r$.

\begin{figure}
\begin{center}
\leavevmode
\includegraphics[scale=0.9]{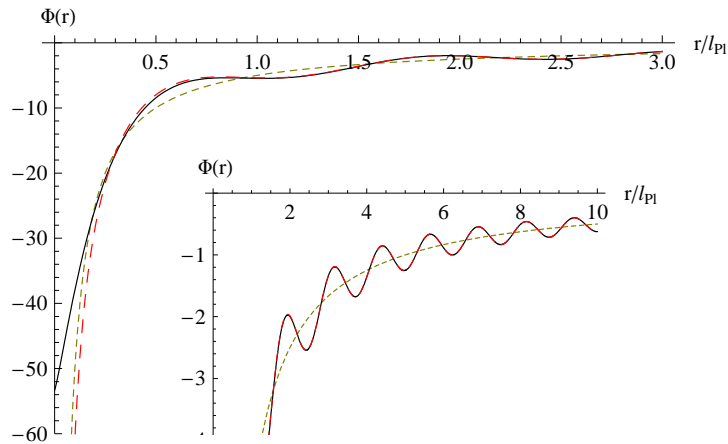}
\end{center}
\caption{The quantum corrected Bardeen potential $\Phi$, obtained
numerically, is represented by the solid line; the long-dashed line
represents its analytical expression~(\ref{Bardeen Phi}) for large
$r$ behavior; the short-dashed line shows Newtonian potential. Values
of the parameters are $\xi=0$ (or $\xi=1/3$), $\tilde \mu=1$ and
$r_S/l_{Pl}=10$.} \label{figPhi}
\end{figure}

\begin{figure}
\begin{center}
\leavevmode
\includegraphics[scale=0.9]{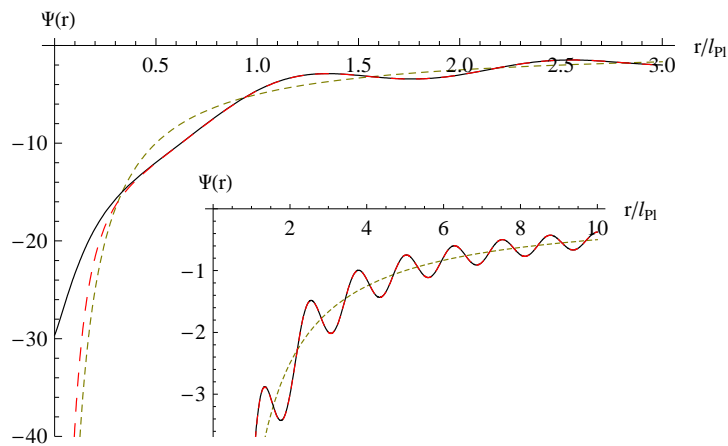}
\end{center}
\caption{The quantum corrected Bardeen potential $\Psi$, obtained
numerically, is represented by the solid line; the long-dashed line
represents its analytical expression~(\ref{Bardeen Psi}) for large
$r$ behavior; the short-dashed line shows Newtonian potential. Values
of the parameters are $\xi=0$ (or $\xi=1/3$), $\tilde \mu=1$ and
$r_S/l_{Pl}=10$.} \label{figPsi}
\end{figure}

\begin{figure}
\begin{center}
\leavevmode
\includegraphics[scale=0.9]{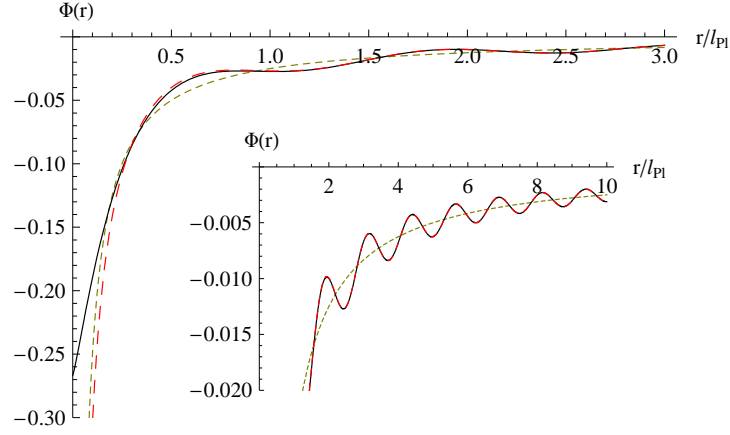}
\end{center}
\caption{The quantum corrected Bardeen potential $\Phi$, obtained
numerically, is represented by the solid line; the long-dashed line
represents its analytical expression~(\ref{Bardeen Phi}) for large
$r$ behavior; the short-dashed line shows Newtonian potential. Values
of the parameters are $\xi=0$ (or $\xi=1/3$), $\tilde \mu=1$ and
$r_S/l_{Pl}=0.05$.} \label{figPhi2}
\end{figure}

\begin{figure}
\begin{center}
\leavevmode
\includegraphics[scale=0.9]{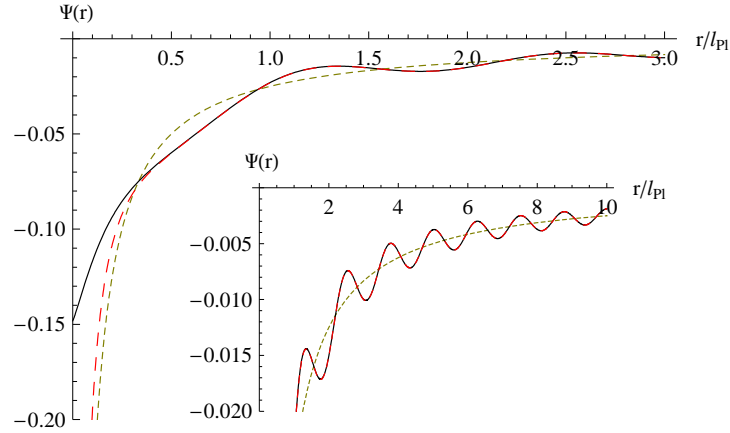}
\end{center}
\caption{The quantum corrected Bardeen potential $\Psi$, obtained
numerically, is represented by the solid line; the long-dashed line
represents its analytical expression~(\ref{Bardeen Psi}) for large
$r$ behavior; the short-dashed line shows Newtonian potential. Values
of the parameters are $\xi=0$ (or $\xi=1/3$), $\tilde \mu=1$ and
$r_S/l_{Pl}=0.05$.} \label{figPsi2}
\end{figure}

For small $r$ and for nonconformal coupling ($\xi\neq 1/6$)
both integrals are proportional to $r$ with the constant of proportionality
depending on the coupling strength $\xi$ only.
This is the main result of this section as it clearly shows that
the $1/r$ singularity is screened on Planck scales by quantum effects
(see Figs.~\ref{figPhi} and~\ref{figPsi} for $r_S/l_{Pl}=10$ and
Figs.~\ref{figPhi2} and~\ref{figPsi2} for $r_S/l_{Pl}=0.05$, where $r_S=2G_N M$ is
the Schwarzschild radius). Nevertheless, in the case of the conformal
coupling ($\xi=1/6$) the second integral in the above expressions
reduces to a constant while the first one is proportional to $r$.
This means that for the conformal coupling there is no screening
of the Newtonian singularity (in this case quantum effects do not
increase the order of singularity at the origin),
which means that quantum effects are the weakest in the conformally coupled case.

For large $r$, the first integral, denoted by $I_1(y)$
\beq
I_1(y)=\int_0^{\infty}\frac{dx \sin(x y)}{x}
\frac{1}{1+\frac{x^2}{120\pi}\ln\left(\frac{x^2}{4\tilde\mu^2}\,
                   {\rm e}^{2\gamma_E}\right)},
\label{I 1a}
\eeq
is rather simple and yields a constant $ I_1(y)={\pi}/{2}$,
while the second one, we denote by $I_2(y)$,
\beq I_2(y)=\int_0^\infty \frac{\sin(x y)dx}{x}
\frac{1}{1-(1-6\xi)^2\frac{x^2}{24\pi}\ln\left(\frac{x^2}{4\tilde\mu^2}\,
                  {\rm e}^{2\gamma_E}\right)}
\label{I 2a}
\eeq
is more delicate and needs to be treated heedfully. As
$\xi\neq 1/6$, the denominator in this integral has poles on
the real axis denoted as $\pm x_c$. In order to calculate the 
principal part of the integral, we approximate the log-function by a
constant evaluated at $x_c$. This is justified by the fact that
the log-function is the largest (but finite) at the pole.
In this approximation the integral becomes

\beq I_2(y)=-\frac{1}{2} \int_0^\infty dx \sin(x y)
\left(\frac{1}{x-x_c}+\frac{1}{x+x_c}-\frac{2}{x}\right)\,,
\label{I 2b}
\eeq
where $x_c=x_c(\xi,\mu)$ is the (positive, real) solution
 of the transcendental equation
\begin{equation}
 x_c^2\times \frac{(1-6\xi)^2}{24\pi}\ln\left(\frac{x_c^2}{4\tilde\mu^2}\,
                  {\rm e}^{2\gamma_E}\right) = 1
\,.
\label{xc}
\end{equation}
The approximate integral~(\ref{I 2b}) can be expressed in terms of
the sine integral $\mbox{Si}(x)$ and the cosine
integral $\mbox{Ci}(x)$, defined by
\beq
\mbox{Si}(x)=-\int_x^\infty\frac{\sin(z) dz}{z}\,,\qquad
\mbox{Ci}(x)=-\int_x^\infty\frac{\cos(z) dz}{z}\,.
\eeq
Using asymptotic expansions of $\mbox{Si}(x)$ and $\mbox{Ci}(x)$,
we get the following approximate expression for
$I_2$ in the limit of large $r$:
\beq
 I_2(y)\approx \frac{\pi}{2 }\left[1- \cos( x_c y)\right]
\,.
\nonumber
 \eeq
 This result needs to be checked against the numerical one.
It turns out that one gets agreement, provided
one modifies the coefficient in front of
the cosine function by a function $a(\xi)$ that weakly depends on $\xi$.
So a better approximation for asymptotically large $r$ is
\beq
 I_2(y)=\frac{\pi}{2 }\left[1- a(\xi)\cos( x_c y)\right]
\,,
 \eeq
where $a(\xi)\sim 1$.\footnote{The precise analytic form of the function
$a(\xi)$ is not known to us
[because we do not know how to perform the integrals~(\ref{I 1a}) and (\ref{I 2a})].
For us it suffices to note that $a(\xi)$ is of the order 1 and it rather
weakly depends on $\xi$ (for conformal coupling $\xi=1/6$, $a=0$ vanishes).
}
When this and $I_1=\pi/2$
are inserted into~(\ref{Phi:4}--\ref{Psi:4}) one arrives
at the following asymptotic form for the Bardeen potentials:
\bea
 \Phi&=&-\frac{G_N M}{r}\bigg\{1+\frac 1 3 a(\xi)\cos(\kappa_c r)\bigg\}
 \label{Bardeen Phi}\\
 \Psi&=&-\frac{G_N M}{r}\bigg\{1-\frac 1 3 a(\xi)\cos(\kappa_c r)\bigg\}
 \,,
\label{Bardeen Psi}
\eea
where $y x_c=\kappa_c r$, $\kappa_c=x_c/l_{Pl}$.
Notice that, unlike in the classical case,
the one-loop scalar quantum fluctuations generate
different Bardeen potentials at large $r$.
The Bardeen potentials exhibit oscillatory behavior at large $r$.
These oscillations occur at the Planck scale, and their amplitude
is about 1/3 of the classical potential
(modulated by a weakly dependent function of $\xi$ of order 1).
When the effective action of Ref~\cite{Satz:2010uu} is used
to calculate quantum corrections to the Newtonian potential,
apart from the usual $\propto 1/r^3$ term, the authors
of~\cite{Satz:2010uu} obtained oscillatory terms
$\propto \sin(\mu r)/r^{5/2}$, where $\mu$ is the renormalization scale.
It would be of interest to find out if there is a relation to the oscillatory
terms found in this work.
Equations~(\ref{Bardeen Phi}) and~(\ref{Bardeen Psi}) comprise the second
main result of this work.

 In Figs.~\ref{figPhi} -- \ref{figPsi2}
we show the numerically evaluated quantum corrected Bardeen
potentials as a function of the distance from the point mass $M$.
While Figs.~\ref{figPhi} and~\ref{figPsi} show the Bardeen potentials for a heavy mass,
$M\gg m_{Pl}$, Figs.~\ref{figPhi2} and~\ref{figPsi2} show the Bardeen potentials induced by
a light mass, $M\ll m_{Pl}$ (which holds for the mass of all known elementary particles).
The Figs. show that, when
quantum fluctuations are included (solid lines), the Newtonian potentials
(short-dashed lines) get antiscreened such that
the Newtonian singularity $\propto 1/r$ gets resolved.
Furthermore, for large $r$, there is very good agreement between our
semianalytical expressions for
the Bardeen potentials~(\ref{Bardeen Phi}) and~(\ref{Bardeen Psi})
and the curves obtained by numerical integration.
Next, we show how the potentials depend on the $r_S/l_{Pl}$ parameter.
In general, if the Schwarzschild radius is greater than
the Planck length then an event horizon forms (a property characterizing
black holes), in which case the Bardeen potential $\Psi$ crosses
$-1/2$. This is exactly what we see in Fig.~\ref{figPsi}. However, if the Schwarzschild radius
is less than the Planck length (as it is the case
with all known elementary particles) then no event horizon forms,
i.e. $\Psi>-1/2$ for all $r$. This feature is clearly manifested
in Figs.~\ref{figPhi2} and~\ref{figPsi2}.

 We have seen that quantum fluctuations antiscreen gravitational potentials
in the sense that they reach a finite value when $r\rightarrow 0$.
For small $r$ the potentials are linear in $r$, i.e.
$\Phi\simeq \Phi_0 + \Phi_0^\prime r$,
$\Psi\simeq \Psi_0 + \Psi_0^\prime r$. The resulting gravitational force,
$\vec F = -\nabla \Phi = -\Phi_0^\prime \vec r/r$ is constant and negative,
implying that the force reaches a maximum amount. This is to be contrasted
with the Newton force, which grows without a limit as $\propto 1/r^2$
at small $r$. Let us now consider the Riemann curvature tensor
$R_{\rho\sigma\mu\nu}$.
When linearized in $h_{\mu\nu}$,
\begin{equation}
 R_{\rho\sigma\mu\nu} = \partial_\mu\Big[\partial_{(\nu}h_{\sigma)\rho}
                        -\frac12\partial_\rho h_{\nu\sigma}\Big]
                        - \partial_\nu\Big[\partial_{(\mu}h_{\sigma)\rho}
                        -\frac12\partial_\rho h_{\mu\sigma}\Big]
\,.
\label{Riemann:linear}
\end{equation}
Taking account of symmetries of the Riemann tensor,
the nonvanishing components of~(\ref{Riemann:linear}) can be written
in terms of gauge invariant components of $h_{\mu\nu}$ only as
\begin{eqnarray}
 R_{0i0j} &=& -R_{i00j} = -R_{i00j} = R_{i0j0} = R_{j0i0} =
             \partial_i\partial_j\Phi
            +\delta_{ij}\partial_0^2 \Psi
            +\partial_0\partial_{(i}\tilde n_{j)}^T
            - \frac12\partial_0^2 h_{ij}^{TT}\,,
\nonumber\\
 R_{0ijl} &=& -R_{i0jl} = R_{jl0i} = -R_{jli0} =
             2\partial_0\partial_{[j}\delta_{l]i}\Psi
            - \partial_0 \partial_{[j}h_{l]i}^{TT}\,,
\nonumber\\
 R_{ijkl} &=& -R_{jikl} = -R_{ijlk} = R_{klij} =
             4\partial_{[i}\delta_{j][k}\partial_{l]}\Psi
             +\big(\partial_j \partial_{[k}h_{l]i}^{TT}
                  - \partial_i \partial_{[k}h_{l]j}^{TT}
               \big)
\,.
\label{Riemann:linear:2}
\end{eqnarray}
When applied to the problem at hand ($h_{ij}^{TT}=0$, $\partial_0\Psi=0$,
$\tilde n_i^T=0$), these reduce to
\begin{eqnarray}
 R_{0i0j} &=& \Big(\frac{\delta_{ij}}{r} - \frac{x_ix_j}{r^3}\Big)\Phi_0^\prime
\,,\qquad R_{0ijl} = 0\,,
\nonumber\\
 R_{ijkl} &=& 4\Big(\frac{\delta_{j[k}\delta_{l]i}}{r}
                     -\frac{x_{[i}\delta_{j][k}x_{l]}}{r^3}\Big)\Psi_0^\prime
\,,
\label{Riemann:linear:3}
\end{eqnarray}
from which we can obtain the Ricci curvature tensor and scalar,
\begin{equation}
 R_{00} = \frac{2}{r}\Phi_0^\prime
\,,\qquad
 R_{ij} = -\Big(\frac{\delta_{ij}}{r} - \frac{x_ix_j}{r^3}\Big)\Phi_0^\prime
          -\Big(\frac{3\delta_{ij}}{r} - \frac{x_ix_j}{r^3}\Big)\Psi_0^\prime
\,,\qquad
 R = - \frac{4}{r}\Phi_0^\prime
          -\frac{8}{r}\Psi_0^\prime
\,.
\label{Ricci:small r}
\end{equation}
From these equations we see that a curvature singularity remains.
It is however reduced
from the standard curvature singularity of a black hole,
$R_{\rho\sigma\mu\nu}R^{\rho\sigma\mu\nu} =  (48 G_N^2M^2)/r^6$
(which is present only in the Weyl part of the Riemann tensor since
$R_{\mu\nu}=0$) to much milder curvature-squared singularities,
\begin{equation}
 R_{\mu\nu} R^{\mu\nu}
     =\frac{2}{r^2}(3{\Phi_0^\prime}^2 + 6\Psi_0^\prime\Phi_0^\prime
                          +11{\Psi_0^\prime}^2)
\,,\qquad
 R_{\rho\sigma\mu\nu} R^{\rho\sigma\mu\nu} =\frac{8}{r^2}(\Phi_0^{\prime\,2}+3\Psi_0^{\prime\,2})\,,
\end{equation}
which for small $r$ are of the order $\sim 1/r^2$. It would be of interest to check whether this behavior
persists at two-loop order, or perhaps the screening of gravitational potentials is more efficient
when higher loop effects are taken into account. Such an analysis would increase our understanding of
the strength of gravity at very small (Planckian) scales.

\section{Quantum gravitational radiation}
\label{Quantum gravitational radiation}

 Here we discuss how the one-loop vacuum polarization from
 (nonminimally coupled massless) scalars~(\ref{retarded self energy})
 affects the propagation of gravitational waves. For simplicity,
 we perform here a perturbative analysis of the
 1PI resummed equation~(\ref{A:hijTT:2}), and we shall work in direct space.
This section is inspired by the recent paper of Leonard and Woodard~\cite{Leonard:2012fs},
where it was shown that the photon one-loop vacuum polarization induced by the graviton vacuum fluctuations
affects propagation of the photons by deforming its light cone.
Here we shall see that the analogous effect is present in the case of
dynamical gravitons immersed in a sea of vacuum fluctuations of matter fields.

We begin our analysis by a simple quadrupole model, for which nonvanishing components
of the conserved energy momentum tensor are
\begin{eqnarray}
 T_{00}(\vec x,t) &=& \tilde I_{ij} r(t) \partial_i\partial_j\delta^3(\vec x)
 \nonumber\\
 T_{0i}(\vec x,t) &=& \tilde I_{ij} \Theta(t)\partial_j\delta^3(\vec x)
 \nonumber\\
 T_{ij}(\vec x,t) &=& \tilde I_{ij} \delta(t)\delta^3(\vec x)
 \,,
\label{stress energy: quadrupole}
\end{eqnarray}
where $\tilde I_{ij} = \int d^ 4x {\rm e}^{\imath k\cdot x} T_{ij}(x)$
is the transverse, traceless, momentum space moment of inertia, satisfying $\tilde I_{ii}=0$ and
$\partial_i \tilde I_{ij}=0$ and $r(t)$ is a "ramp" function $dr(t)/dt=\Theta(t)$, and
$d\Theta(t)/dt=\delta(t)$. Of course, the stress energy tensor~(\ref{stress energy: quadrupole})
is an approximation to
a realistic explosive event with a nonvanishing quadrupole, which has a finite
spatial extension. The classical, gauge invariant equation for
the (transverse-traceless) graviton is
\begin{equation}
 \partial^2 h_{ij}^{(c)TT}(x) = -16\pi G_N T_{ij}
 \,
\label{hij:classical}
\end{equation}
and its solution for the stress energy tensor~(\ref{stress energy: quadrupole}) is
the Einstein quadrupole formula,
\begin{equation}
 h_{ij}^{(c)TT}(x) = 4G_N \tilde I_{ij}\frac{\delta(t-r)}{r}
                    =  8G_N \tilde I_{ij}\Theta(t)\delta(x^2)
 \,,\qquad x^2 = -t^2+r^2,\quad r \equiv \|\vec x\|
\,.
\label{tensor: classical solution}
\end{equation}
An inspection of Eq.~(\ref{A:hijTT:2}) shows that, at this order, quantum effects do not mix
tensor perturbations with scalar or vector perturbations.
Thus, Eqs.~(\ref{A:quantum in terms of cl}) and (\ref{A:Frhosigma})
can be rewritten for $h_{ij}^{TT}$ as
\bea
h^{(c)TT}_{ij}(x)=h^{TT}_{ij}(x)-\frac{\kappa^2}{7680\pi^3}
\bigg\{D_{ijkl}-\delta_{ij}D_{kl}
  \bigg[\frac{1}{6}+30\Big(\xi-\frac{1}{6}\Big)^2\bigg]\bigg\}F_{kl}(x)
\,,
\label{quantum in terms of cl:tensor}
 \eea
where
\begin{equation}
F_{kl}(x) = \partial^2\int
d^4 x^\prime\Theta(\Delta t^2-\Delta r^2)\Theta(\Delta
t)\left[\ln(\mu^2(\Delta t^2-\Delta
r^2))-1\right]{\partial^\prime}^2 h_{kl}^{TT}(x^\prime)
\,,
\label{Fkl}
\end{equation}
where we have moved by partial integration two out of four derivatives to act on $h_{kl}^{TT}$.
Equations~(\ref{quantum in terms of cl:tensor}) and (\ref{Fkl}) can be solved perturbatively, by
writing $h_{kl}^{TT}= h_{kl}^{(c)TT} + h_{kl}^{(q)TT}$, where  $h_{kl}^{(c)TT}$
is the classical solution~(\ref{tensor: classical solution}) and $h_{kl}^{(q)TT}$ is
the quantum correction. Inserting this
into~(\ref{quantum in terms of cl:tensor}) and (\ref{Fkl}) results in
\begin{equation}
h^{(q)TT}_{ij}(x)=\frac{G_N}{480\pi^2}
\bigg\{D_{ijkl}-\delta_{ij}D_{kl}
  \bigg[\frac{1}{6}+30\Big(\xi-\frac{1}{6}\Big)^2\bigg]\bigg\}F_{kl}(x)\,.
\label{quantum in terms of cl:tensor:2}
\end{equation}
Inserting $h_{kl}^{(c)TT}$ into~(\ref{Fkl})
yields the leading order (one-loop) quantum correction to
the graviton field. Making use of~(\ref{hij:classical}), Eq.~(\ref{Fkl})
simplifies to
\begin{equation}
F_{kl}(x) = -16\pi G_N\tilde I_{ij}
    \partial^2\left\{\Theta(t^2-r^2)\Theta(t)\left[\ln(\mu^2(t^2-r^2))-1\right]
              \right\}
\,.
\label{Fkl:2}
\end{equation}
Acting naively with the last two derivatives is problematic as it leads to
singular expressions. This  can be resolved by recalling that in the
Schwinger-Keldysh formalism the retarded vacuum polarization originates
from
\begin{eqnarray}
F_{kl}(x) &=& -16\pi G_N\tilde I_{ij}
    \partial^2\left\{\frac{\left[\ln^2(\mu^2 x^2_{++})-2\ln(\mu^2x^2_{++})\right]
                    -\left[\ln^2(\mu^2 x^2_{+-})-2\ln(\mu^2x^2_{+-})\right]}{4\pi\imath}
              \right\}
\nonumber\\
     &=& -\frac{32 G_N}{\imath}\tilde I_{kl}
    \left\{\frac{\ln(\mu^2 x^2_{++})}{x^2_{++}}
                    -\frac{\ln(\mu^2 x^2_{+-})}{x^2_{+-}}
              \right\}
\,,
\label{Fkl:3}
\end{eqnarray}
where $x_{++}^2=-(|t|-\imath\epsilon)^2+r^2$, $x_{+-}^2=-(t+\imath\epsilon)^2+r^2$.
As a final step we insert this result into
Eq.~(\ref{quantum in terms of cl:tensor:2}). Two important observations
are in order. Firstly, the operator $D_{ij}$ defined in Eq.~(\ref{D1:D2}) yields zero
when it acts on a traceless, transverse tensor, implying that
the term in~(\ref{quantum in terms of cl:tensor})
containing the Ricci-scalar coupling $\xi$ does not contribute
to the graviton propagation, i.e. a minimally and nonminimally coupled
scalar contribute equally to graviton propagation. Secondly,
the operator $D_{ijkl}$ defined in Eq.~(\ref{D1:D2}) yields a very simple
result when it acts on a transverse, traceless tensor,
 $D_{ijkl}h_{kl}^{TT}=-(1/2)\partial^2h_{ij}^{TT}$. With these observations,
 it is now easy to evaluate the leading quantum correction to the graviton
field~(\ref{quantum in terms of cl:tensor:2}). The result is
\begin{equation}
h^{(q)TT}_{ij}(x)=-\frac{2G_N^2}{15\pi^2\imath}
\tilde I_{ij}\left\{\frac{1}{x^4_{++}}
                    -\frac{1}{x^4_{+-}}
              \right\}
               = \frac{2G_N^2}{15\pi^2\imath}
\tilde I_{ij}\frac{\partial}{\partial r^2}\left\{\frac{1}{x^2_{++}}
                    -\frac{1}{x^2_{+-}}
              \right\}
= -\frac{4G_N^2}{15\pi}\tilde I_{ij}\frac{\partial}{\partial r^2}\Theta(t)\Theta(t^2-r^2)
\,.
\label{quantum in terms of cl:tensor:3}
\end{equation}
Combining this with the classical graviton field~(\ref{tensor: classical solution}) gives
\begin{equation}
 h_{ij}^{TT}(x) = 8G_N \tilde I_{ij}\left(1-\frac{G_N}{30\pi}\frac{\partial}{\partial r^2}\right)
                 \theta(t)\delta(t^2-r^2)+{\cal O}(G_N^3)
                =  8G_N \tilde I_{ij}\delta\Big(t^2-r^2+\frac{G_N}{30\pi}\Big)+{\cal O}(G_N^3)\,.
\label{quantum in terms of cl:tensor:4}
\end{equation}

Just as in Ref.~\cite{Leonard:2012fs}, where the authors found that the photons in presence of
graviton quantum fluctuations move slightly superluminally,
we have found here that the gravitons in presence of scalar quantum fluctuations
move also slightly superluminally,\footnote{If one defines
the graviton speed as the ratio $r/t$, then Eq.~(\ref{shifted lightcone}) implies that
gravitons move slightly superluminally. However, if one looks at the local
speed $dr/dt = t/r$ (which is more in the spirit of group velocity), then the local
speed is subluminal, and as $t\rightarrow 0$ it even reaches zero. This is of course just a
curiosity, since the light cone deformation~(\ref{shifted lightcone}) becomes perturbatively
trustable only for super-Planckian times, $t\gg t_{\rm pl}$.} 
i.e., they move on a shifted light cone given by
\begin{equation}
  r^2 = t^2 + \frac{G_N}{30\pi}
\,.
\label{shifted lightcone}
\end{equation}
\begin{figure}
\begin{center}
\leavevmode
\includegraphics[scale=0.8]{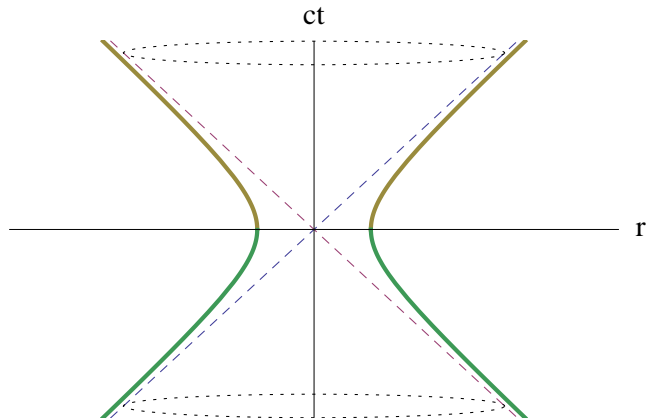}
\end{center}
\caption{The classical and quantum corrected light cones of
dynamical gravitons.} \label{figLightCone}
\end{figure}

Curiously, the effect does not depend on $\xi$, i.e. it does not depend on whether
the scalar is minimally or nonminimally coupled. This means that the effect~(\ref{shifted lightcone})
comes entirely from the kinetic term cubic vertex $-h^{\mu\nu}(\partial_\mu\varphi)(\partial_\nu\varphi)/2$.
The classical light cones, as well as the quantum-shifted light cones, are
shown in Fig.~\ref{figLightCone}.
Just as in the case of the photons, this graviton effect is very tiny (in fact, it is 40 times smaller than
the effect found for the photons in~\cite{Leonard:2012fs}), and thus it is uncertain
whether it will ever be observed. From the structure of the light cone in Fig.~\ref{figLightCone},
it is clear that the superluminality is not cumulative (it does not increase with distance).
On the contrary, it becomes weaker as the distance increases. This can be understood as follows: the
effect on larger scales is built primarily from quantum fluctuations on that scale, which
have smaller amplitude than fluctuations on a smaller scale, and vice versa
Yet, the effect is of some academic interest,
as it is for the first time that a superluminal propagation of gravitons is
inferred from quantum matter fluctuations. As a curiosity, we mention that
a similar deformation of light cones has been observed in Hermitean Gravity
of Ref.~\cite{Mantz:2008hm}, where a slight superluminal behavior was observed for particles
with negative 4-momentum squared, $-p^2=[(E/c)^2-\vec p^{\,2}]>0$. Our effect is also different
from the superluminal photon motion observed in curved
geometries~\cite{Drummond:1979pp,Daniels:1995yw,Daniels:1993yi},
where the effects were calculated in a gradient expansion. No such approximation has been used here.
Moreover, our results are manifestly gauge independent,
and they are calculated within a gauge invariant formalism.

\section{Summary and discussion}
\label{Summary and discussion}

 In this paper we perform one-particle irreducible (1PI) resummation
of the one-loop vacuum fluctuations of nonminimally coupled, massless,
scalar matter.
%and study its (backreaction) effects on classical point
%particles moving on bound orbits.
Our main result~(\ref{Bardeen Phi}) and (\ref{Bardeen Psi})
indicates that gravity gets strongly antiscreened on
(super-)Planckian energies,
supporting the hypothesis of asymptotic freedom, according to which
gravity becomes weak on super-Planckian scales. Furthermore, we study
how vacuum fluctuations affect graviton propagation.
We find that gravitons moving in a sea of vacuum fluctuations are sped up,
resulting in a slightly superluminal motion (see Fig.~\ref{figLightCone}). However, the effect is too weak to be
observable.

 Antiscreening on short scales is such that the Bardeen
gravitational potentials remain
finite everywhere, also arbitrarily close to the origin. However,
the Riemann curvature tensor still exhibits singularity at the origin,
albeit this singularity is much milder ($R_{\mu\nu\rho\sigma}R^{\mu\nu\rho\sigma}\propto 1/r^2$)
than the usual singularity of black hole space times (for which
$R_{\mu\nu\rho\sigma}R^{\mu\nu\rho\sigma}\propto 1/r^6$).
In addition, the Ricci scalar and Ricci tensor are also singular, which is not surprising
since vacuum fluctuations act as matter.
Since for (elementary) particles of sub-Planckian mass
the Bardeen potentials remain small everywhere (see Figs.~\ref{figPhi2} and~\ref{figPsi2}),
our results may be trustable when applied to known elementary particles.
For heavy particles (whose mass is
comparable to or larger than the Planck mass), we also find qualitatively
the same strong antiscreening.
However, our results cannot be trusted here since,
at short distances (closer to the would-be event horizon),
the amplitude of the gravitational potentials becomes of the order 1 or
larger (see Figs.~\ref{figPhi} and~\ref{figPsi}), invalidating our linearized approach to gravitational perturbations.
So, in order to obtain more reliable results also in this case,
it is required to extend our methods beyond linear order in the gravitational
fields, or work in coordinates in which perturbations remain small everywhere.
But, this task is left for a future work.

\medskip

 A couple of remarks follow. It is believed that, just like quantum chromodynamics (QCD),
 quantum gravity is an asymptotically free theory, in the
sense that in the far ultraviolet it decouples from matter fields and becomes weak.
There is some evidence for the asymptotic safety program that
in the UV quantum gravity develops correlations
as if it were effectively a two-dimensional theory. This then implies
that the corresponding spectrum is scale invariant, which may
have relevance for generation of cosmological perturbations.
Asymptotic freedom of QCD has led to the conjecture (which has
in the meantime been confirmed by numerical simulations)
that, at high temperatures, the QCD matter is in a state of a quark-gluon
plasma, in which most of the quarks (those of a sufficiently high energy)
move essentially free.
A quite exciting possibility is that a similar scenario may be realized within
gravity. One can then imagine a collection of gravitationally weakly interacting very
closely packed particles which would make dense objects
which -- due to the gravitational asymptotic freedom -- would not collapse
into black holes, or at least would screen curvature singularities at the
origin of black holes. Whether this picture is consistent with
our understanding of quantum gravity remains to be seen
and requires further investigation.

Our analysis in the Appendix contains results that are worth commenting on.
The graviton vacuum polarization tensor
of a massless nonminimally coupled scalar field has such a tensor structure
that, when it acts on a graviton field, it yields gauge invariant field components.
This result motivates the following {\it conjecture}:

\begin{enumerate}
\item[]
Gauge invariance dictates the structure of the graviton vacuum polarization tensor
induced by any massless matter or gravitational field at the one-loop order.
The structure is a slight generalization of~(\ref{retarded self energy}),
\begin{equation}
\left[{}_{\mu\nu}\Pi_{\rho\sigma}^{\rm ret}\right](x;x^\prime)
  = L_{\mu\nu\alpha\beta}
\bigg\{\alpha D^{\alpha\beta}{}_{\rho\sigma}+\beta\eta^{\alpha\beta}D_{\rho\sigma}
 \bigg\}
 \partial^4\, \Big\{\Theta(\Delta t^2-\Delta r^2)\Theta(\Delta t)
  \left[\ln| \mu^2(\Delta r^2-\Delta t^2)|-1\right]\Big\}\,,
\,
\nonumber
\end{equation}
where $\alpha$ and $\beta$ are field dependent constants
and $D^{\rho\sigma}$ and $D_{\alpha\beta}^{\;\;\;\;\rho\sigma}$
are the derivative operators defined in~(\ref{D1:D2})
 [if a field is massive, then the scalar function
on the second line of~(\ref{retarded self energy}) will be different,
but on the light cone it will reduce to the one given
in~(\ref{retarded self energy}) and in the equation above].
The vacuum polarization tensor at higher loops will still have the same
tensor structure, but scalar functions will generally be different.
\end{enumerate}

Conversely, the structure of Eq.~(\ref{Full EOM}) is such that it is
gauge invariant.
Since this equation is linear in metric perturbations,
it also means that it is gauge independent to the (linear) order
considered. This can be shown as follows. Firstly, note that both
the graviton vacuum polarization~(\ref{retarded self energy})
and the Lichnerowicz operator~(\ref{Lichnerowicz})
consist of operators $D_{\mu\nu}$ and $D_{\mu\nu\rho\sigma}$ defined in
Eqs.~(\ref{D1:D2}). Now, since under an infinitesimal coordinate shift,
$x^\mu\rightarrow x^\mu + \xi^\mu(x)$, where $\xi^\mu(x)$ is an infinitesimal
function, to linear order the metric perturbation
$h_{\mu\nu}$ transforms as $h_{\mu\nu}\rightarrow h_{\mu\nu}-2\partial_{(\mu}\xi_{\nu)}$.
This then implies that, to linear order in $\xi^\mu$ we have
\begin{equation}
 D^{\mu\nu} h_{\mu\nu}\rightarrow D^{\mu\nu} h_{\mu\nu}
           - 2 D^{\mu\nu} \partial_{\mu}\xi_{\nu}
           = D^{\mu\nu} h_{\mu\nu}\,,
\nonumber
\end{equation}
and
\begin{equation}
 D^{\mu\nu\rho\sigma} h_{\rho\sigma}\rightarrow D^{\mu\nu\rho\sigma} h_{\rho\sigma}
           - 2 D^{\mu\nu\rho\sigma} \partial_{\rho}\xi_{\sigma}
           = D^{\mu\nu\rho\sigma} h_{\rho\sigma}
\,.
\nonumber
\end{equation}
This immediately implies that the classical part of Eq.~(\ref{Full EOM})
is both gauge independent and gauge invariant.
In the quantum part of~(\ref{Full EOM}) one can perform
two partial integrations to move $D^{\mu\nu}$ and $D^{\mu\nu\rho\sigma}$ to act
on $h_{\rho\sigma}$ under the $x^\prime$ integral, implying that
-- up to boundary terms -- the whole equation can be written in
a gauge invariant form.
The invariance under some symmetry transformation of
the vacuum polarization tensor
can be checked by considering, for example, the vacuum polarization induced
by the one-loop (Abelian or non-Abelian)
gauge fields~\cite{Capper:1979ej,Capper:1978yf,Capper:1974vb}.
An intricate gauge dependence in the one-loop gravitational potentials
seen by a point particle was discussed in
Refs.~\cite{Dalvit:1997yc,Gribouk:2003ks}. It would be of interest
to relate these results and to the ones presented in this work.

A fully covariant form
for the vacuum polarization was presented in~\cite{Jordan:1987wd}
and in~\cite{Barvinsky:1987uw,Barvinsky:1990up}.
Next we discuss how to connect our work to the covariant
perturbation theory of Barvinsky and
Vilkovisky~\cite{Barvinsky:1987uw,Barvinsky:1990up}.
Adapting Eq.~(3.4) from~\cite{Barvinsky:1990up}
to a nonminimally coupled scalar field in a Lorentzian space yields for
the one-loop effective action
\begin{eqnarray}
 \Gamma_{\rm eff} &=& -\frac{1}{\kappa^2}\int d^D x \sqrt{-g} R
\label{BarvinskyVilkoviskyAction}\\
&&\hskip -0cm
    -\,\frac{1}{1920\pi^2} \int d^D x \sqrt{-g}
\bigg\{ R_{\alpha\beta}\Big[\Gamma(-\Box)
   \!+\!\frac{16}{15}\,\Big]
   R^{\alpha\beta}\! + 30\Big(\xi\!-\!\frac16\Big)^2
        R\Gamma(-\Box)R
    -\frac13R\Gamma(-\Box)R
    -\frac{10}{3}\Big(\xi\!-\!\frac{29}{100}\Big)R^2
\bigg\}
\,,
\quad
\nonumber
\end{eqnarray}
where we included the Einstein-Hilbert classical action
and the finite part of the operator $\Gamma(-\Box)$ is
\begin{equation}
 \Gamma(-\Box) = \!-\!\ln\Big(\frac{-\Box}{4\pi\mu^2}\Big)
               + 2 - \gamma_E
\,.
\label{BV:nonlocal operator}
\end{equation}
In Eq.~(\ref{BarvinskyVilkoviskyAction}) we took the values for the tensors
from Eq.~(3.4) that correspond to a nonminimally coupled scalar field,
that is $\hat P = [(1/6) - \xi]R\hat 1$ and
${\cal R}_{\mu\nu}=0$, and we evaluated the trace (${\rm tr} [\hat 1] = 1$).
Note that in an Euclidean space the operator $\ln(-\Box/\mu^2)$
in Eq.~(3.4) of Ref.~\cite{Barvinsky:1990up} is well defined
because the norm of the operator $-\Box$ is positive. This is however
not the case in a Lorentzian space and hence extra care is needed to
properly (uniquely) define the meaning of the nonlocal operator
$\ln(-\Box/\mu^2)$. The correspondence between the Euclidean result
(\ref{BV:nonlocal operator}) and the Lorentzian {\tt in-in} formulation
(and in particular the representation
of the nonlocal operator $\ln(-\Box/\mu^2)$ in the {\tt in-in} formalism)
can be found, for example, in Ref.~\cite{Lombardo:1996gp}.

 Next we observe that, to quadratic order in metric perturbation $h_{\mu\nu}$
around flat Minkowski space, the action~(\ref{BarvinskyVilkoviskyAction})
becomes,
\begin{eqnarray}
 \Gamma_{\rm eff} &=& \frac{1}{2\kappa^2}\int d^D x
            h^{\mu\nu}L_{\mu\nu\rho\sigma} h^{\rho\sigma}
\label{BarvinskyVilkoviskyAction:quadratic}\\
&&\hskip -0cm
    -\,\frac{1}{1920\pi^2} \int d^D x h^{\mu\nu}
\bigg\{ \Big[\Gamma(-\Box)\!+\!\frac{16}{15}\,\Big]
     L_{\mu\nu\alpha\beta}
\Big(
D^{\alpha\beta}_{\;\;\rho\sigma}-\frac12\eta^{\alpha\beta}D_{\rho\sigma}
\Big)
\nonumber\\
&&\hskip -0cm
 +\, \bigg[30\Big(\xi\!-\!\frac16\Big)^2-\frac13\bigg]
  \Gamma(-\Box)L_{\mu\nu\alpha\beta}\Big(-\eta^{\alpha\beta}D_{\rho\sigma}\Big)
    -\frac{10}{3}\Big(\xi\!-\!\frac{29}{100}\Big)
  L_{\mu\nu\alpha\beta}\Big(-\eta^{\alpha\beta}D_{\rho\sigma}\Big)
\bigg\}h^{\rho\sigma}
\,,
\quad
\nonumber
\end{eqnarray}
where we made use of $R=D_{\mu\nu}h^{\mu\nu}+{\cal O}(h_{\alpha\beta}^2)$,
$R_{\alpha\beta}=D_{\alpha\beta\mu\nu}h^{\mu\nu}+{\cal O}(h_{\alpha\beta}^2)$
and $D_{\mu\nu}$ and $D_{\alpha\beta\mu\nu}$ are defined in
Eq.~(\ref{D1:D2}) and, in addition, we partially integrated twice
and made use of Eqs.~(25) and (26) from Ref.~\cite{Marunovic:2011zw}.
Now, upon varying the action~(\ref{BarvinskyVilkoviskyAction:quadratic})
and setting it equal to the stress energy tensor of a point mass source
$(1/2)\delta_\mu^{\;0}\delta_\nu^{\;0}M\delta^3(\vec x\,)$
we get,
\begin{eqnarray}
h^{\alpha\beta\,(c)} =  h^{\alpha\beta}
    \!+\frac{\kappa^2}{960\pi^2}
\bigg\{ \bigg[\!-\!\Gamma(-\Box)\!-\!\frac{16}{15}\,\bigg]
D^{\alpha\beta}_{\;\;\rho\sigma}
 - \bigg[\frac16\!+\!30\Big(\xi\!-\!\frac16\Big)^2\bigg]
  \big[\!-\!\Gamma(-\Box)\big]\eta^{\alpha\beta}D_{\rho\sigma}
   \!-\!\frac{10}{3}\Big(\xi\!-\!\frac{9}{20}\Big)\eta^{\alpha\beta}D_{\rho\sigma}
\bigg\}h^{\rho\sigma}
,
\quad
\label{BarvinskyVilkoviskyAction:quadratic:2}
\end{eqnarray}
where we made use of the fact that $h^{\rho\sigma\,(c)}$ is the classical
metric that corresponds to the pointlike source
[see Eq.~(\ref{EOM cl})],
which allowed us to remove the Lichnerowicz operator
from Eq.~(\ref{BarvinskyVilkoviskyAction:quadratic:2}).
When the (real space)
expression~(\ref{BarvinskyVilkoviskyAction:quadratic:2}) is compared
with the momentum space expression~(\ref{A:quantum vs classical:2})
we see that there is perfect agreement of the coefficients
multiplying the logarithm
$\ln(-\partial^2/\mu^2)\rightarrow \ln(|(k^0)^2-k^2|/\mu^2) +$
(an imaginary part), while the coefficients of the local terms do not agree.
These coefficients are not universal and can be adjusted
by a suitable choice of local counterterms.
\footnote{In fact, these coefficients of local terms
in~(\ref{BarvinskyVilkoviskyAction:quadratic:2})
are not universal in the
Barvinsky-Vilkovisky covariant perturbation theory,
because the relation between the metric perturbation and the corresponding
curvature tensor perturbations is not gauge independent, and thus can vary
depending on how one fixes the gauge, as can be seen
from Eq.~(3.9) of Ref.~\cite{Barvinsky:1987uw}.}
A generalization of the Barvinsky-Vilkovisky results
to massive field theories has been exacted in
Refs.~\cite{Avramidi:1989er,Dalvit:1994gf} which, of course,
in the massless limit reproduce
Eq.~(\ref{BarvinskyVilkoviskyAction:quadratic:2}).
In conclusion, we
have shown that our one-loop expression for the metric perturbations
around Minkowski background can be obtained from
the more general expression derived in
Refs.~\cite{Barvinsky:1987uw,Barvinsky:1990up}.
As regards the imaginary part of the operator~(\ref{BV:nonlocal operator}),
we note that the prescription used in the
Appendix corresponds to the retarded boundary prescription,
which is the appropriate (causal) prescription that automatically follows from
the Schwinger-Keldysh formalism. An alternative prescription for the imaginary
part corresponding to the retarded self-energy was given
in Ref.~\cite{Jordan:1987wd}, where the author argued that
the graviton self-energy can be represented in terms of the
Feynman propagators.

Finally, it would be of interest to clarify what is the connection (if any)
between the results of the asymptotic safety program
(according to which quantum gravity in the ultraviolet becomes weak, and hence antiscreened)
and the results presented in this work.

\section*{Appendix: Solving the integral equation for the gravitational field}
\label{Appendix: Solving the integral equation for the gravitational field}

 Here we show how to solve the integral equation~(\ref{quantum in terms of cl})
for the (quantum) gravitational field $h^{\alpha\beta}(x)$,
\bea
h_{\alpha\beta}^{(c)}(x)&=&h_{\alpha\beta}(x)-\frac{\kappa^2}{7680\pi^3}
\bigg\{D_{\alpha\beta}{}^{\rho\sigma}-\eta_{\alpha\beta}D^{\rho\sigma}
  \bigg[\frac{1}{6}+30\Big(\xi-\frac{1}{6}\Big)^2\bigg]\bigg\}F_{\rho\sigma}(x)
\,,
\label{A:quantum in terms of cl}
 \eea
where
\begin{equation}
F_{\rho\sigma}(x) = \partial^4\int
d^4 x^\prime\Theta(\Delta t^2-\Delta r^2)\Theta(\Delta
t)\left[\ln(\mu^2(\Delta t^2-\Delta
r^2))-1\right]h_{\rho\sigma}(x^\prime).
\label{A:Frhosigma}
\end{equation}
Upon performing a spatial Fourier transform, Eq.~(\ref{A:Frhosigma})
can be reduced to
\begin{equation}
 F_{\rho\sigma}(\vec k,t) = \frac{4\pi}{k}(\partial_t^2+k^2)^2
  \int_{-\infty}^tdt^\prime\int_0^{\Delta t}d(\Delta r)\Delta r
   \sin(k\Delta r)\Big[\ln\Big(\mu^2(\Delta t^2-\Delta r^2)\Big)-1\Big]
       h_{\rho\sigma}(\vec k,t^\prime)
\,,
\label{A:Frhosigma:2}
\end{equation}
where $k=\|\vec k\|$ and
\begin{equation}
 F_{\rho\sigma}(\vec k,t)=\int d^3 x
            {\rm e}^{-\imath \vec k\cdot \vec x}
                  F^{\rho\sigma}(\vec x,t)
\,,\qquad
 h_{\rho\sigma}(\vec k,t^\prime)=\int d^3 x
            {\rm e}^{-\imath \vec k\cdot \vec x}
                  h_{\rho\sigma}(\vec x,t^\prime)
\,.
\nonumber
\end{equation}
The spatial integral can be reduced to $\Xi_{-1}(z,\zeta)$
defined in Ref.~\cite{Prokopec:2003iu}, Eqs.~(45)-(50),
\begin{equation}
 F_{\rho\sigma}(\vec k,t) = \frac{4\pi}{k^3}(\partial_t^2+k^2)^2
  \int_{-\infty}^tdt^\prime \Xi_{-1}(z,\zeta) h_{\rho\sigma}(\vec k,t^\prime)
  \,,
\label{A:Frhosigma:3}
\end{equation}
where $z=k\Delta t$, $\zeta = k/\mu$, $x=\Delta r/\Delta t$,
\begin{equation}
\Xi_n(z,\zeta) = z^2\int_0^1 dx x\sin(zx)[2\ln(z/\zeta)+\ln(1-x^2)+n]
               = [\sin(z)-z\cos(z)][2\ln(z/\zeta)+n]+z^2\xi(z)
\,
\label{A:Xin}
\end{equation}
and
\begin{eqnarray}
\xi(z) &=& z^2\int_0^1 dx x\sin(zx)\ln(1-x^2)
\nonumber\\
               &=& 2\sin(z) -[\cos(z)+z\sin(z)][{\rm si}(2z)+\pi/2]
                + [\sin(z)-z\cos(z)][{\rm ci}(2z)-\gamma_E-\ln(z/2)]
\,,
\label{A:xi}
\end{eqnarray}
with
\begin{equation}
{\rm si}(z) = -\int_z^\infty\frac{\sin(t)dt}{t}
            =\int_0^z\frac{\sin(t)dt}{t} -\frac{\pi}{2}
\,,\qquad
{\rm ci}(z) = -\int_z^\infty\frac{\cos(t)dt}{t}
            =\int_0^z\frac{[\cos(t)-1]dt}{t} +\gamma_E+\ln(z)
\,.
\nonumber
\end{equation}
Now, since for small $z=k\Delta t$, $\Xi_n\propto z^3\ln(z)$, one can
act with $\partial_t^2+k^2$ once without touching the upper limit of the
time integral in~(\ref{A:Frhosigma:3}) to obtain
\begin{equation}
 F_{\rho\sigma}(\vec k,t) = \frac{4\pi}{k}(\partial_t^2+k^2)
  \int_{-\infty}^tdt^\prime (\partial_z^2+1)\Xi_{-1}(z,\zeta)
          h_{\rho\sigma}(\vec k,t^\prime)
\,,
\label{A:Frhosigma:4}
\end{equation}
where
\begin{equation}
(\partial_z^2+1)\Xi_{n}(z,\zeta)
 = \sin(z)[2\ln(z/\zeta) +(n+1)+{\rm ci}(2z)-\gamma_E-\ln(2z)]
   -\cos(z)[{\rm si}(2z)+\pi/2]
\nonumber
\end{equation}
Next, upon observing that,
\begin{equation}
(\partial_z^2+1)^2\Xi_{n}(z,\zeta)
 = 4\frac{\cos(z)-1}{z} +4\partial_z [\ln(2z/\zeta)+(n+1)/2]
\,,
\nonumber
\end{equation}
one concludes that~(\ref{A:Frhosigma:4}) can be recast as
\begin{equation}
 F_{\rho\sigma}(\vec k,t) = 16\pi\Bigg[
  \int_{-\infty}^tdt^\prime \frac{\cos(k\Delta t)-1}{\Delta t}
   +\partial_t \int_{-\infty}^tdt^\prime \ln(2\mu\Delta t)\Bigg]
          h_{\rho\sigma}(\vec k,t^\prime)
\,.
\label{A:Frhosigma:5}
\end{equation}
We shall now argue that in the second integral, one can move
the derivative $\partial_t$ to act as $\partial_{t^\prime}$ on
$h_{\rho\sigma}(\vec k,t^\prime)$.
Notice firstly that one can write
 $\ln(2\mu\Delta t)=-\partial_{t^\prime}\{ \Delta t[\ln(2\mu\Delta t)-1]\}$,
and then one can partially integrate $\partial_{t^\prime}$
to act on $h_{\rho\sigma}(\vec k,t^\prime)$.
Formally, this will generate an infinite contribution from the initial state
at $t^\prime \rightarrow -\infty$. This infinite contribution can be
regulated away, if one assumes that
$h_{\rho\sigma}(\vec k,t^\prime)$ gets adiabatically switched on as
$h_{\rho\sigma}(\vec k,t^\prime)\propto {\rm e}^{\epsilon t^\prime}$
as $t^\prime \rightarrow -\infty$. After this, one can act with
$\partial_t$, and the second integral becomes
$\int_{-\infty}^t dt^\prime \ln(2\mu\Delta t)
           \partial_{t^\prime}h_{\rho\sigma}(\vec k,t^\prime)$.
This expression is now prepared for a Fourier transform
over time. By noting that
\begin{equation}
 h_{\rho\sigma}(\vec k,t^\prime)
   = \int \frac{dk^0}{2\pi}{\rm e}^{-\imath k^0t^\prime} h_{\rho\sigma}(k^\mu)
  = {\rm e}^{-\imath k^0t} \int \frac{dk^0}{2\pi} {\rm e}^{\imath k^0\Delta t}
    h_{\rho\sigma}(k^\mu)
\,,
\nonumber
\end{equation}
$F_{\rho\sigma}(k^\mu)$ can be written in momentum space as,
\begin{equation}
F_{\rho\sigma}(k^\mu) = -8\pi
  \bigg[
       \ln\bigg(\frac{|(k^0)^2-k^2|}{4\mu^2}\bigg)+2\gamma_E
          -\frac{\imath\pi}{2}\Big({\rm sign}(k^0+k)+{\rm sign}(k^0-k)\Big)
  \bigg]h_{\rho\sigma}(k^\mu)
\,,
\label{A:Frhosigma:6}
\end{equation}
where we made use of the following integrals,
\begin{eqnarray}
\int_0^\infty d\Delta t {\rm e}^{\imath k^0\Delta t}
                   \frac{\cos(k\Delta t)-1}{\Delta t}
   &=& \frac12\ln\bigg(\frac{|k^0|^2}{|(k^0)^2-k^2|}\bigg)
 + \frac{\imath \pi}{4}\Big[{\rm sign}(k^0+k)+{\rm sign}(k^0-k)-2{\rm sign}(k^0)\Big]\,,
\nonumber\\
\int_0^\infty d\Delta t {\rm e}^{\imath k^0\Delta t}\ln(2\mu\Delta t)
   &=& -\frac{\imath}{k^0}
    \bigg[
        \ln\bigg(\frac{|k^0|}{2\mu}\bigg)
            -\frac{\imath \pi}{2}{\rm sign}(k^0)+\gamma_E
     \bigg]
\,.
\nonumber
\end{eqnarray}
Now making use of
$(1/2)[{\rm sign}(k^0+k)+{\rm sign}(k^0-k)]={\rm sign}(k^0)\theta((k^0)^2-k^2)$
we see that~(\ref{A:Frhosigma:6}) can be also written as
\begin{equation}
F_{\rho\sigma}(k^\mu) = -8\pi
  \bigg[
       \ln\bigg(\frac{-(k^0+\imath\epsilon)^2+k^2}{4\mu^2}\bigg)+2\gamma_E
  \bigg]h_{\rho\sigma}(k^\mu)
\,.
\label{A:Frhosigma:6b}
\end{equation}
This structure of the retarded one-loop self-energy agrees
with  e.g. Eq.~(E9) in Appendix~E of Ref.~\cite{Koksma:2009wa},
where various components of the one-loop self-energy within the
Schwinger-Keldysh {\it in-in} formalism are
constructed for an interacting scalar field theory.
 It is not clear to us how to relate~(\ref{A:Frhosigma:6b}) to the
prescription  used in~\cite{Jordan:1987wd}, where it was argued that
the retarded self-energy is obtained by taking the real part of
the contribution
$\propto\ln[(-(k^0)^2+k^2-\imath\epsilon)/\mu^2]
=\ln\big[|(k^0)^2-k^2|/\mu^2\big]-\imath\pi\theta\big((k^0)^2-k^2\big)$,
whereby the real, causal part should be taken after the momentum
integration.\footnote{The result~(2.19) in Ref.~\cite{Jordan:1987wd} is
sloppy in that no proper care was taken of the fact that
the Heaviside function $\theta(t-t^\prime)$ (used to select for causality)
does not commute with the derivatives in Eq.~(2.18).
The procedure advocated in this work and in~\cite{Koksma:2009wa,Koksma:2011dy}
does not suffer from such problems.}
The result~(\ref{A:Frhosigma:6}) or~(\ref{A:Frhosigma:6b})
for the retarded self-energy is simpler and more
transparent, and we shall therefore use it in the rest of this work.
 Inserting~(\ref{A:Frhosigma:6}) into Eq.~(\ref{A:quantum in terms of cl})
and transforming into momentum space one gets,
\bea
h_{\alpha\beta}^{(c)}(k^\mu)
 &=&h_{\alpha\beta}(k^\mu)+\frac{\kappa^2}{960\pi^2}
\bigg\{\tilde D_{\alpha\beta}{}^{\rho\sigma}-\eta_{\alpha\beta}\tilde D^{\rho\sigma}
  \bigg[\frac{1}{6}+30\Big(\xi-\frac{1}{6}\Big)^2\bigg]\bigg\}
  \nonumber \\
    && \times\,  \bigg[
       \ln\bigg(\frac{|(k^0)^2-k^2|}{4\mu^2}\bigg)+2\gamma_E
          -\frac{\imath\pi}{2}\Big({\rm sign}(k^0+k)+{\rm sign}(k^0-k)\Big)
  \bigg]h_{\rho\sigma}(k^\mu)
\,,
\label{A:quantum vs classical:2}
 \eea
where [{\it cf.} Eq.~(\ref{D1:D2})]
\beq
\tilde D^{\rho\sigma}=-k^{\rho}k^\sigma+\eta^{\rho\sigma}
          (k^{\mu}k_{\mu})
\;,\qquad
\tilde D_{\alpha\beta}^{\;\;\;\;\rho\sigma} =
-k_{(\alpha}\delta_{\beta)}^{\;\;(\rho}k^{\sigma)}
 + \frac{1}{2}\delta_{\alpha}^{\;\;(\rho}\delta^{\sigma)}_{\;\;\beta}(k^{\mu}k_{\mu})
         + \frac{1}{2}\eta^{\rho\sigma}k_\alpha k_\beta
\,.
\label{A:D1:D2}
\eeq
Next it is useful to act with the two operators~(\ref{A:D1:D2}) on $h_{\rho\sigma}$.
It is interesting to note that these generate only gauge invariant combinations
of components of $h_{\rho\sigma}$. For example,
\begin{equation}
\tilde D^{\rho\sigma}h_{\rho\sigma}(k^\mu)
  = 2 k^2\Phi(k^\mu) + 2\big[3(k^0)^2-2k^2\big]\Psi(k^\mu)\,,
\label{A:Drs hrs}
\end{equation}
where
\begin{equation}
\Psi =-\frac16(h+k^2 \tilde h)
\,,\qquad
\Phi =-\frac12(h_{00}+2\imath k^0\sigma -(k^0)^2 \tilde h)\,,
\label{A:Bardeen potentials}
\end{equation}
and we have made use of a Helmholtz decomposition of $h_{\mu\nu}$,
\begin{equation}
h_{0i}= n_i^T+\imath k_i\sigma
\,,\qquad \
h_{ij}=\frac{\delta_{ij}}{3}h -\Big(k_ik_j-\frac{\delta_{ij}}{3}k^2\Big)\tilde h
       +\imath (k_i h_j^T+k_j h_i^T) + h_{ij}^{TT}
\label{A:Helmholtz decomposition}
\end{equation}
where $n_i^T,h_i^T$ and $h_{ij}^{TT}$ are transverse,
$k^i n_i^T = 0=k^i h_i^T$, $k^ih_{ij}^{TT} = 0 = h_{ij}^{TT}k^j$, and
$h_{ij}^{TT}$ is in addition traceless, $h_{ii}^{TT}=0$ and
$k\equiv \|\vec k\|$.
Apart from the Bardeen potentials, there is one gauge invariant vector and
one gauge invariant tensor,
\begin{equation}
  \tilde n_i^T = n_i^T + \imath k^0h_i^T
\,,\qquad h_{ij}^{TT}
\,,
\label{A:gi vector and tensor}
\end{equation}
such that there are in total six gauge invariant fields, which means
that four out of the ten components of $h_{\mu\nu}$ are gauge dependent,
i.e. they change if one performs a coordinate shift
$x^\mu\rightarrow x^\mu+\xi^\mu(x)$,
where $\xi^\mu(x)$ is a small coordinate shift,
of the order of metric components $h_{\mu\nu}$.
Note that $h_{ij}^{TT}$ is on its own gauge invariant, and represents the two
dynamical, tensorial degrees of freedom of gravity (gravitational waves).

 Similarly, the action of the second operator in~(\ref{A:D1:D2})
yields gauge invariant quantities only,
\begin{eqnarray}
\tilde D_{00}^{\;\;\;\;\rho\sigma}h_{\rho\sigma}
   &=& - k^2 \Phi - 3(k^0)^2\Phi\,,
\nonumber\\
\tilde D_{0i}^{\;\;\;\;\rho\sigma}h_{\rho\sigma}
   &=& \frac12 k^2 \tilde n_i^T +2 k^0 k_i\Psi
    = \tilde D_{i0}^{\;\;\;\;\rho\sigma}h_{\rho\sigma}\,,
\nonumber\\
\tilde D_{ij}^{\;\;\;\;\rho\sigma}h_{\rho\sigma}
   &=&  k_i k_j\Phi +\big[-k_ik_j-\delta_{ij}\big(-(k^0)^2+k^2\big)\big]\Psi
         -k^0k_{(i} \tilde n_{j)}^T +\frac12\big[-(k^0)^2+ k^2\big]h_{ij}^{TT}
\,.
\label{A:D2:action}
\end{eqnarray}
Since the Lichnerowicz operator~(\ref{Lichnerowicz})
of the classical equation of motion for $h_{\rho\sigma}^{(c)}$
is a linear combination of $D^{\rho\sigma}$ and
$D_{\alpha\beta}^{\;\;\;\;\rho\sigma}$, it also yields gauge invariant quantities
in the equation of motion. This observation was already made in
our earlier work~\cite{Marunovic:2011zw}. There is however a much stronger statement
one can make: the structure of the equation that includes the
graviton vacuum polarization~(\ref{Full EOM}) and (\ref{retarded self energy})
is also gauge invariant (at least to the order the equation is written).
This property can be used to our advantage and study quantum effects
in a fully gauge invariant fashion.

 Since all components of the vacuum polarization are gauge invariant,
one can use the Helmholtz decomposition~(\ref{A:Helmholtz decomposition})
and simply combine the component equations~(\ref{A:quantum vs classical:2})
into gauge invariant combinations according to~(\ref{A:Bardeen potentials})
and~(\ref{A:gi vector and tensor}), resulting in the following
gauge invariant equations:
\bea
\Phi^{(c)}&=&\Phi+{\cal K}
      \Big\{
        \big(-(k^0)^2+k^2\big)\Phi+\Big[\frac16+30\Big(\xi-\frac16\Big)^2\Big]
           \Big[-2k^2\Phi+2\big(-3(k^0)^2+2k^2\big)\Psi\Big]
      \Big\}\,,
\label{A:Phi}\\
\Psi^{(c)}&=&\Psi+{\cal K}
      \Big\{
        \big(-(k^0)^2+k^2\big)\Psi-\Big[\frac16+30\Big(\xi-\frac16\Big)^2\Big]
           \Big[-2k^2\Phi+2\big(-3(k^0)^2+2k^2\big)\Psi\Big]
      \Big\}\,,
\label{A:Psi}\\
{\tilde n}_i^{T(c)}&=&\tilde n_i^T+{\cal K}
      \big(-(k^0)^2+k^2\big)\tilde n_i^T
\label{A:ni}\\
{h_{ij}^{TT}}^{(c)}&=&h_{ij}^{TT}+{\cal K}
      \big(-(k^0)^2+k^2\big)h_{ij}^{TT}
\,,
\label{A:hijTT}
 \eea
where we defined
\begin{equation}
{\cal K} = \frac{\kappa^2}{1920\pi^2}\bigg[
       \ln\bigg(\frac{|(k^0)^2-k^2|}{4\mu^2}\bigg)+2\gamma_E
          -\frac{\imath\pi}{2}\Big({\rm sign}(k^0+k)+{\rm sign}(k^0-k)\Big)
  \bigg]
\,.
\label{A:cal K}
\end{equation}
The last two equations~(\ref{A:ni}) and (\ref{A:hijTT}) are easily solved,
\bea
{\tilde n}_i^{T}(k^\mu)
       &=&\frac{{\tilde n}_i^{T(c)}(k^\mu)}{1+\big(-(k^0)^2+k^2\big){\cal K}(k^\mu)}\,,
\label{A:ni:2}\\
{h_{ij}^{TT}}(k^\mu)
   &=&\frac{{h_{ij}^{TT}}^{(c)}(k^\mu)}{1+{\cal K}(k^\mu)\big(-(k^0)^2+k^2\big)}
\,,
\label{A:hijTT:2}
 \eea
and tell us that the effect of one-loop vacuum polarization
for the vector and tensor perturbations can be absorbed into a finite
wave function renormalization; i.e. vacuum polarization
changes the amplitude of the gravitational wave. The effect is however
tiny, since on shell, where $-(k^0)^2+k^2=0$, the effect vanishes.
That means that, when there are no classical gravitational waves,
quantum effects of scalar vacuum fluctuations do not generate any.
The same holds for vector perturbations:
if they are zero classically, they will not be generated by
fluctuating scalar fields. Because the structure of vacuum polarization
is dictated by gauge invariant tensor structures, we expect the same
conclusions concerning gauge invariant vector and tensor perturbations
to hold for arbitrary fields: the quantum effects of the fluctuations
will just renormalize the wave function.

 The structure of the scalar equations~(\ref{A:Phi}) and (\ref{A:Psi})
is more complex, because they in general couple.
Note that the equation for the sum of the Bardeen potentials is particularly simple,
\begin{equation}
\Phi^{(c)}+\Psi^{(c)}=\Phi+\Psi + {\cal K}\big(-(k^0)^2+k^2\big)(\Phi+\Psi)
\,,
\label{A:Phi+Psi}
 \end{equation}
whose solution is of the same form as that of
the vector~(\ref{A:ni:2}) and tensor perturbations~(\ref{A:hijTT:2}),
\begin{equation}
\Phi+\Psi=\frac{\Phi^{(c)}+\Psi^{(c)}}{1 + {\cal K}\big(-(k^0)^2+k^2\big)}
\,;
\label{A:Phi+Psi:solution}
 \end{equation}
i.e. the effect of loop fluctuations of quantum fields
on the sum $\Phi+\Psi$ is just a finite renormalization of its amplitude,
which vanishes on shell.

 It is, of course, possible to decouple the two Bardeen
potentials~(\ref{A:Phi}) and (\ref{A:Psi}). A convenient way of writing
the decoupled equations is
\begin{eqnarray}
\Phi &=& \frac{\Phi^{(c)}}{1-5(1-6\xi)^2(k^\mu k_\mu){\cal K}}
         -\Big(\frac23k^2-(k^0)^2\Big){\cal K}
             \Bigg[\frac{5(1-6\xi)^2}{1-5(1-6\xi)^2(k^\mu k_\mu){\cal K}}
                +\frac{1}{1+(k^\mu k_\mu){\cal K}}
             \Bigg](\Phi^{(c)}+\Psi^{(c)})\,,
\qquad
\label{A:Phi:2}\\
\Psi &=& \frac{\Psi^{(c)}}{1-5(1-6\xi)^2(k^\mu k_\mu){\cal K}}
         -\frac13k^2{\cal K}
             \Bigg[\frac{5(1-6\xi)^2}{1-5(1-6\xi)^2(k^\mu k_\mu){\cal K}}
                +\frac{1}{1+(k^\mu k_\mu){\cal K}}
             \Bigg](\Phi^{(c)}+\Psi^{(c)})
\,.
\label{A:Psi:2}
 \end{eqnarray}
This means that each of the two quantum-corrected Bardeen potentials
depend on both classical Bardeen potentials. Since in general relativity,
the classical potentials sourced by a point mass (or a scalar field)
are equal $\Phi^{(c)} = \Psi^{(c)}$, Eqs.~(\ref{A:Phi:2}) and (\ref{A:Psi:2})
further simplify to
\begin{eqnarray}
\Phi &=& \Phi^{(c)}\Bigg[-\frac13\frac{1-10(1-6\xi)^2(k^0)^2{\cal K}}
                                      {1-5(1-6\xi)^2(k^\mu k_\mu){\cal K}}
                +\frac23\frac{2+(k^0)^2{\cal K}}{1+(k^\mu k_\mu){\cal K}}
             \Bigg]\,,
\qquad
\label{A:Phi:3}\\
\Psi &=& \Phi^{(c)}\Bigg[\frac13\frac{2-5(1-6\xi)^2(k^0)^2{\cal K}}
                                      {1-5(1-6\xi)^2(k^\mu k_\mu){\cal K}}
                +\frac13\frac{1-(k^0)^2{\cal K}}{1+(k^\mu k_\mu){\cal K}}
             \Bigg]
\,.
\label{A:Psi:3}
 \end{eqnarray}
For a point stationary mass we have $\Phi^{(c)}=\Psi^{(c)}= 4\pi G_NM/k^2$.
These equations are now used to obtain the static limit of
a (one-loop) quantum corrected gravitational response to a point mass
in the main text.

\end{document}